\documentclass[12pt,letterpaper]{article}
\usepackage{amsmath}
\usepackage{fancyhdr}
\usepackage{graphicx}
\usepackage[left=2cm,top=2cm,right=2cm]{geometry}
\usepackage[small]{caption}

\DeclareMathAlphabet{\mathpzc}{OT1}{pzc}{m}{it}

\begin{document}

\begin{center}
\begin{Large}
\bf{Optical detection of single non-absorbing molecules using the surface plasmon of a gold nanorod}\\
\end{Large}
\vskip 3 cm {Peter Zijlstra$^1$, Pedro M.R. Paulo$^2$, and Michel Orrit$^{+1}$} \\
\vskip 1 cm
$^1$ MoNOS Huygens Laboratorium, Universiteit Leiden, 2300 RA Leiden, The
Netherlands  \vskip 0.3 cm
$^2$ Centro de Qu\'imica Estrutural -- Complexo I, Instituto Superior T\'ecnico, \\ Av. Rovisco Pais, 1049-001 Lisboa, Portugal \vskip 1 cm
$^+$e-mail: orrit@physics.leidenuniv.nl \vskip 3 cm
compiled \today
\end{center}

\clearpage

\textbf{Introductory paragraph}
Current optical detection schemes for single molecules require light absorption, either to produce fluorescence \cite{moerner99} or direct absorption signals \cite{gaiduk10,chong10,kukura10}. This severely limits the range of molecules that can be detected, because most molecules are purely refractive. Metal nanoparticles \cite{anker08,mayer11} or dielectric resonators \cite{arnold03,shopova11,lu11} detect non-absorbing molecules by a resonance shift in response to a local perturbation of the refractive index, but neither has reached single-protein sensitivity. The most sensitive plasmon sensors to date detect single molecules only when the plasmon shift is amplified by a highly polarizable label \cite{sonnichsen05a,sannomiya08} or by a localized precipitation reaction on the particle's surface \cite{chen11}. Without amplification, the sensitivity only allows for the statistical detection of single molecules \cite{mayer10}. Here we demonstrate plasmonic detection of single molecules in realtime, without the need for labeling or amplification. We monitor the plasmon resonance of a single gold nanorod with a sensitive photothermal assay \cite{boyer02} and achieve a $\sim$700-fold increase in sensitivity compared to state-of-the-art plasmon sensors \cite{nusz09}. We find that the sensitivity of the sensor is intrinsically limited due to spectral diffusion of the SPR. We believe this is the first optical technique that detects single molecules purely by their refractive index, without any need for photon absorption by the molecule. The small size, bio-compatibility and straightforward surface chemistry of gold nanorods may open the way to the selective and local detection of purely refractive proteins in live cells.

\clearpage

\begin{figure}[ht!]
\begin{center}
\includegraphics{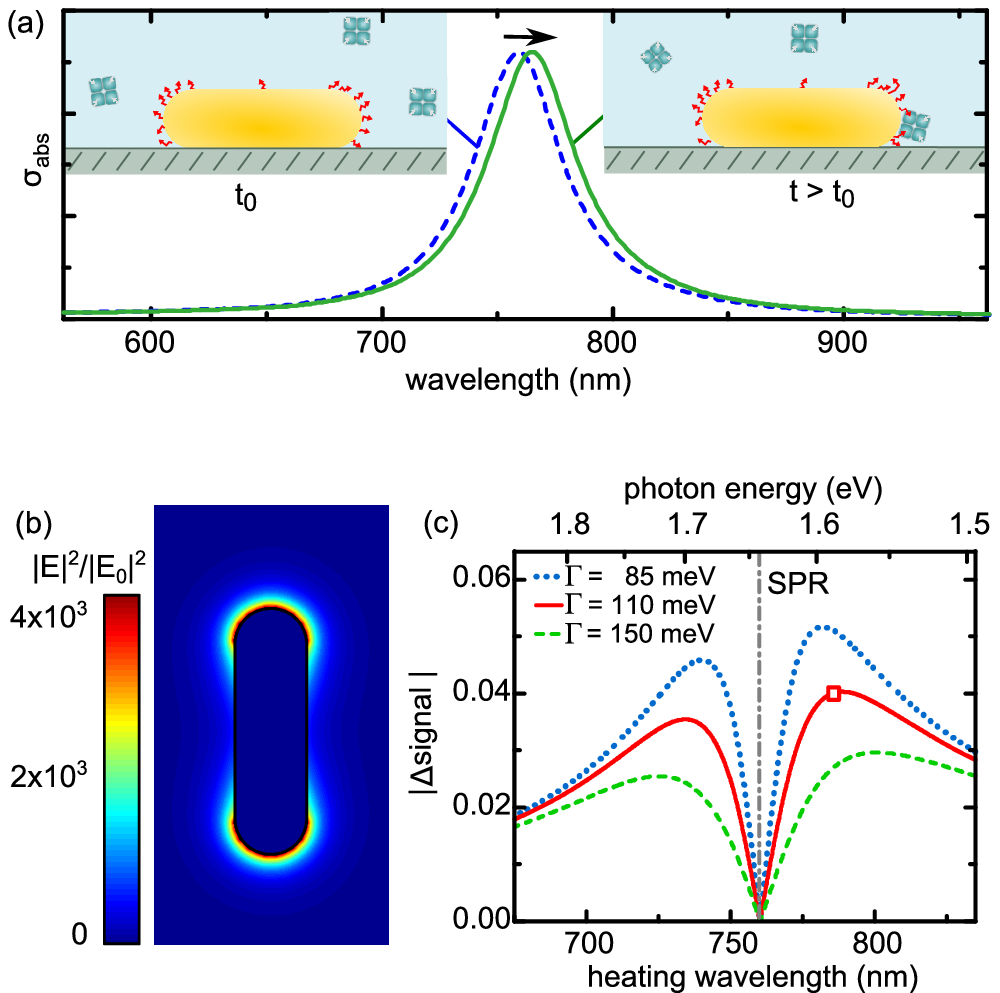}
\caption{\label{principle}}
\end{center}
\end{figure}

\textbf{Main text}
The electric field associated with a surface plasmon resonance (SPR) decays rapidly into the surrounding medium and acts as a transducer that converts changes in the local refractive index into a frequency shift of the SPR. These shifts can be conveniently monitored in the optical far-field without the need for near-field optics or electrical contacts. Sensors based on propagating plasmons are now a mature technology, but they are not suitable to detect single molecules because of their extended area. Single-molecule sensitivity can be attained using surface-enhanced Raman scattering (SERS), which exploits the strong interaction of a molecule with a metal nanostructure, an asperity on a metal surface or a small gap between metal nanoparticles \cite{nie97,kneipp97,xu99,stiles08}. However, SERS-active sites are rare, ill-controlled, and the strong interaction perturbs molecules even more than labels do \cite{etchegoin08}. In contrast, metal nanoparticles offer a lower but reproducible field enhancement \cite{liu07}, and are easily produced in large quantities using wet-chemical synthesis \cite{jana05,zijlstra06}. The principle of our metal particle sensor is schematically shown in Fig. \ref{principle}. The sensor consists of a single gold nanorod (ensemble average size of 37 nm $\times$ 9 nm) that was functionalized with biotin specifically at the tip of the particle (supplementary information section 1). The binding of single molecules to the biotin receptors on the surface of the rod was detected by monitoring the longitudinal SPR at a single frequency using photothermal microscopy.

Photothermal microscopy \cite{boyer02} relies on the intensity change of a detection beam (693 nm) caused by a time-dependent thermal lens (supplementary information section 2). This thermal lens is created by a modulated heating beam of a different color (785 nm) that is absorbed by the nanorod. This technique is capable of detecting very small gold nanoparticles down to 1.4 nm in diameter with a high signal-to-noise ratio (S/N) \cite{berciaud04}, or even single absorbing molecules \cite{gaiduk10}. Unlike the scattering signal, the photothermal contrast scales as the volume of the particle, allowing us to significantly reduce the particle size to obtain a mode-volume that is of the same order as the volume of a typical protein. When a single protein binds to the receptors on the surface of the nanorod, the longitudinal SPR shifts to the red due to the locally increased index of refraction. This shift results in a change of the absorption cross section of the nanorod at the wavelength of the heating beam. The ensuing temperature change is measured by the detection beam as a step-wise change of the photothermal signal.

The magnitude of these steps depends on the strength of the interaction between the SPR and the protein, given by (1) the refractive index contrast between the buffer (n~$\approx$~1.335) and the protein (typically n~=~1.45 for a hydrated protein \cite{voros04}) and (2) the local electric field intensity averaged over the volume of the molecule. The calculated electric field intensity for the longitudinal SPR of a nanorod is shown in Fig. \ref{principle}b. The field is largest at the tips and reduces along the long edges of the particle. By selectively functionalizing the tips of the nanorod we fully exploit the large field enhancement in that region. Moreover, this approach reduces the number of molecules binding to the side of the rod, which are harder to detect individually due to the smaller field intensity.

In Fig \ref{principle}c we show the relative change of the photothermal signal for an SPR red-shift of 1 nm, calculated for different plasmon linewidths. The red-shift is accompanied by a slight increase of the absorption cross section due to the dispersion of the dielectric constant of gold in this wavelength range. As a result, the step size is maximized when the heating laser wavelength is at the half-maximum on the red wing of the longitudinal SPR. Because our heating laser has a fixed wavelength of 785 nm, we selected individual nanorods with a longitudinal SPR at 760 $\pm$ 5 nm by recording the white-light spectrum of the particle before every experiment (Supplementary section 2). The linewidth of the selected rods was typically 110 $\pm$ 5 meV, leading to an estimated 4\% increase of the photothermal signal for a red-shift of 1 nm. In the supplementary information in section 2 we provide an experimental verification of Fig. \ref{principle}c.

\begin{figure}[ht!]
\begin{center}
\includegraphics{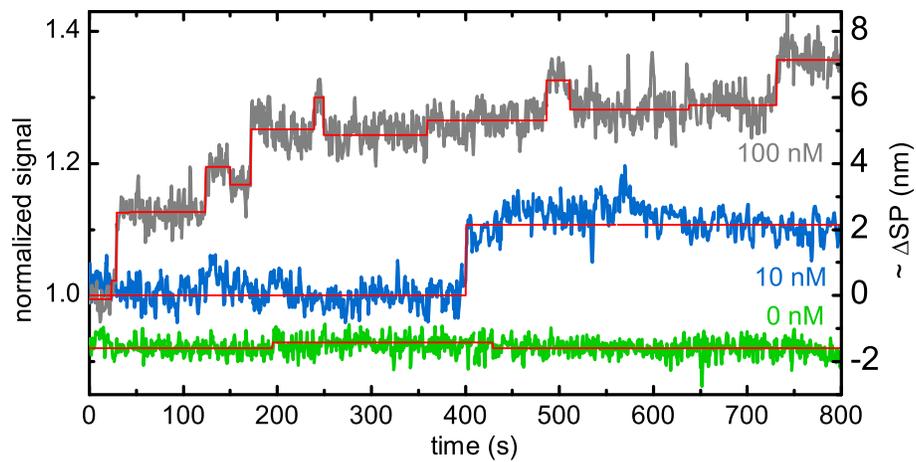}
\caption{\label{timetraces} \textbf{Photothermal timetraces}}
\end{center}
\end{figure}

After identification of a particle that has an appropriate longitudinal SPR, we recorded time traces of the photothermal signal with an integration time of 100 ms (supplementary information section 2). In Fig. \ref{timetraces} we show time traces of 3 different nanorods in the presence of different concentrations of a streptavidin-R-phycoerythrin conjugate (molecular weight 300 kDa). For every nanorod, the photothermal signal was related to a plasmon shift by considering a $\sim$4\% increase of the signal for a 1 nm red-shift (see Fig. \ref{principle}). The actual value slightly varied from nanorod to nanorod, depending on the linewidth as fitted from the white-light spectrum. We attribute the observed increase of the photothermal signal to a red-shift of the longitudinal SPR, which was confirmed by white-light spectra that we recorded immediately before and after each experiment (supplementary information section 3). More importantly, we observe clear steps caused by the binding of individual proteins to biotin groups on the rod surface. The time traces also exhibit a rich collection of dynamic behavior that was not resolvable with previous methods. We observe reversible events, in which the protein leaves the nanoparticle after a few or sometimes several tens of seconds (see for example the case of 100 nM in Fig. \ref{timetraces} at t~=~240 s, 500 s, and 600 s). We hypothesize that these events are due to individual proteins that explore the (tip) surface of the nanorod in search of available biotin. Some proteins may not have the right orientation and conformation to bind a biotin and subsequently desorb and diffuse away.

\begin{figure}[ht!]
\begin{center}
\includegraphics{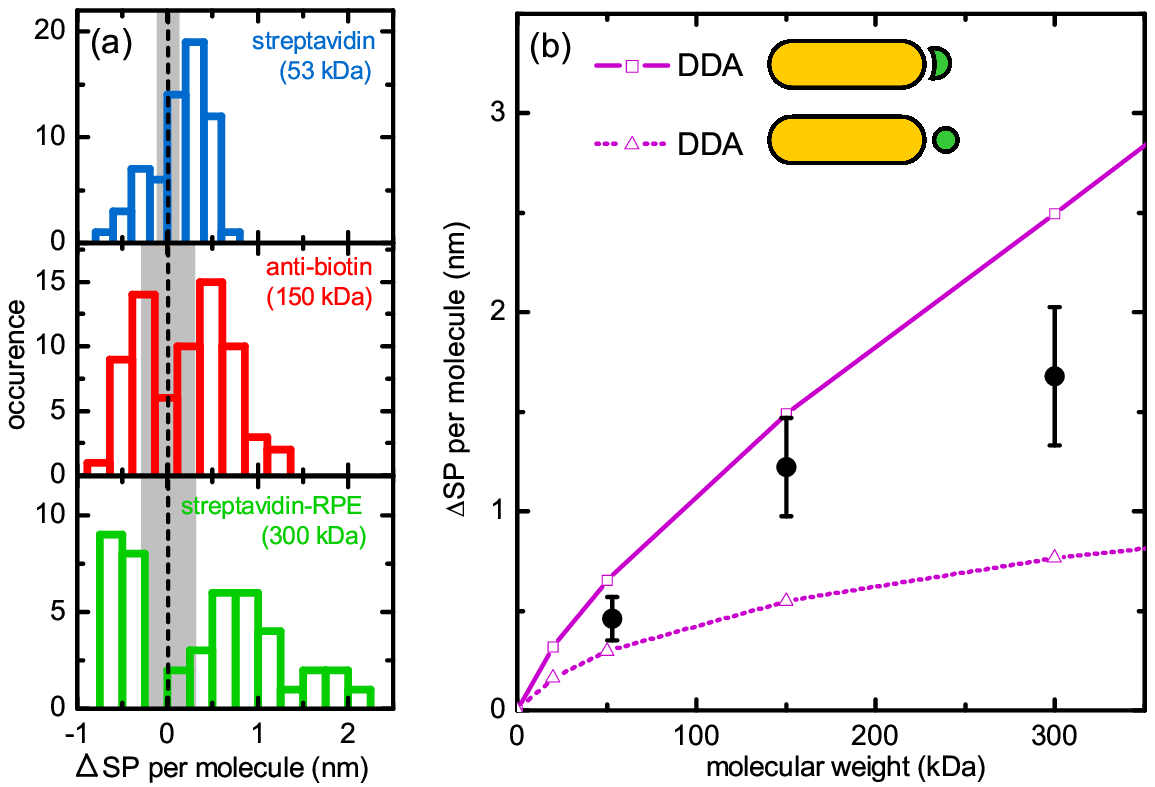}
\caption{\label{statistics}}
\end{center}
\end{figure}

To demonstrate the applicability of our sensor to a range of proteins, we performed experiments on biotin-binding proteins with different molecular weights, namely streptavidin (53 kDa), anti-biotin (150 kDa), and a streptavidin-R-phycoerythrin conjugate (300 kDa). Considering an equal index of refraction for each protein, the heavier proteins are expected to induce a larger step size due to their larger volume. For each protein we acquired several time traces (supplementary information section 3) on different samples, and each trace was fitted with a step-finding algorithm \cite{kerssemakers06}. The accumulated statistics of the fitted step sizes is displayed in the histograms of Fig. \ref{statistics}. For all proteins, we observe a broad distribution of step sizes with a maximum shift value. This behavior is expected as each protein binds at a different location on the surface of the nanorod, yielding different interaction strengths with the SPR. A fraction of the molecules may also bind to the sides of the nanorod, where the step size will be smallest. As expected, we observe a maximum step size (for the proteins that absorb at the tip apex) that scales like the molecular weight of the protein, see Fig. \ref{statistics}b. Blank experiments on rods that were functionalized with thiolated tri-ethylene-glycol instead of biotin did not yield a persistent SPR shift, indicating that the binding events we observe are specific (supplementary information section 4).

The expected SPR shift for a protein binding to the tip apex of the nanorod was calculated using the discrete dipole approximation \cite{yurkin11}, see Fig. \ref{statistics}b and supplementary information section 5. The protein was approximated as a non-absorbing body with a refractive index of 1.45 \cite{voros04}. For the conformation of the protein on the surface of the nanorod we considered two limiting cases, (1) a sphere and (2) an indented sphere that resembles an object that adapted to the curvature of the nanorod's tip. The exact conformation depends on the strength of the protein-surface interaction and may vary from molecule to molecule. The spacing between the protein and the nanorod was kept at 0.5 nm, taking into account that part of the biotin-linker (spacer arm $\sim$~1.5 nm) is embedded in the binding pocket of the protein \cite{weber89}. The measured SPR shifts show excellent agreement with the model and fall in between the two limiting cases we considered.

The sensitivity of our photothermal assay exhibits a significant improvement of more than two orders of magnitude over conventional scattering methods (supplementary information section 6). The ultimate detection limit is determined by the noise on the photothermal signal, defined as its standard deviation for heating at the half-maximum of the SPR. For shot-noise limited detection we expect a noise level that is determined by the photon noise of the detection laser, and is independent of the heating laser power. Instead, for higher heating laser powers we observe an increased noise level (supplementary information section 6). We attribute the observed excess noise to spectral diffusion of the SPR, a phenomenon that has not been observed before but becomes detectable due to the high signal-to-noise ratio of our photothermal assay. To confirm this attribution, we recorded photothermal time-traces of the same nanorod, first without any surface functionalization, and later with a dense thiol coating. The chemical interface damping induced by the thiol coating broadened the SPR of the nanorod from 120 meV to 180 meV, thus reducing the sensitivity to plasmon shifts by $\sim$40\% (see Fig. \ref{principle}c). Indeed, the thiolated nanorod exhibited a noise level that was almost 35\% lower than the uncoated particle. The presence of spectral diffusion suggests an upper bound to the sensitivity of particle plasmon sensors, and currently limits the sensitivity of our method to proteins $>$53 kDa. Further research should establish the influence of the size, shape, and surface chemistry of the particle, which will aid in minimizing the spectral diffusion and maximizing the sensitivity.

The temperature of the excited particle is crucial in these experiments as the structure and activity of the protein can be impaired when it is heated for extended periods of time. Streptavidin has a midpoint for thermal denaturation $T_m$ of 75$^\circ$C \cite{gonzalez99}, and most globular proteins exhibit a $T_m$ ranging from 40$^\circ$C to 80$^\circ$C depending on pH and buffer conditions \cite{robertson97}. From heat conduction calculations we estimate an average temperature rise of $<$2$^\circ$C for our experimental conditions (supplementary information section 7). In the case of streptavidin we employed a higher heating power (to compensate for the smaller step size) resulting in a temperature rise of $\sim$8$^\circ$C. These absolute particle temperatures of less than 25$^\circ$C are far from $T_m$ for the majority of proteins and warrant application of our technique to a wide range of molecules.

Our sensor provides the unique ability to detect purely refractive molecules, without the need for photon absorption by the analyte. Colloidal gold nanorods can be purchased commercially or synthesized in bulk with excellent control over size, aspect ratio and field enhancement. We anticipate that the combination of high reproducibility, straightforward surface chemistry, and nanometer dimensions will spark interest across a range of disciplines. The bio-compatibility combined with the all-optical detection in the far-field makes these nano-sensors particularly useful in biophysical problems. These advances may enable exciting prospects such as label-free detection of analytes inside cells. 
\\
\\
\textbf{Methods summary}\\
\textbf{Sample preparation} The gold nanorods used in this study were prepared by a wet chemical method involving seed-mediated growth in the presence of silver \cite{nikoobakht03}. This yielded single-crystalline nanoparticles with an average length of 37 $\pm$ 4 nm and an average width of 9 $\pm$ 1 nm. After purification, the particles were spincoated onto a (3-mercaptopropyl)trimethoxysilane (MPTS, Sigma) functionalized coverslip. The nanorods were biotinylated in a home-made flow cell by firstly incubating in a 2 mM solution of cetyltrimethylammonium bromide (CTAB) for 30 minutes. This establishes a bilayer of CTAB on the sides of the rod, preventing the thiolated biotin from binding there. The flow cell was then incubated in a 2 $\mu$M buffered solution of thiolated biotin (spacer arm 1.5 nm) in the presence of 2 mM CTAB, for 1 hour. This yielded particles specifically biotinylated at the tips (supplementary information section 1). Before performing single-molecule sensing experiments we characterized the bulk response of our sensor and compared the results to literature values (supplementary information section 8 and 9).\\
\textbf{Optical experiments} For photothermal microscopy, the heating laser beam (Toptica Photonics iBeam Smart, $\lambda$~=~785 nm, modulated at $\Omega$~=~98.3 kHz) was combined with the detection laser beam (Coherent 890 CW titanium sapphire, $\lambda$~=~693 nm). The beams were focused onto the sample through a 1.45 NA oil immersion objective. The reflected detection laser light was detected by a Silicon PIN diode (Femto DHPCA-100-F) attached to a lock-in amplifier (Stanford Research Systems SR844). Before recording timetraces, we minimized the effects of mechanical drift by waiting some time to establish temperature stabilization. All timetraces were recorded with a lock-in integration time of 100 ms. \\
White-light spectra were recorded by strongly focusing a spatially filtered beam from a fiber-coupled xenon lamp (Ocean Optics HPX-2000). The reflected light was directed to a photon counting avalanche photodiode (Perkin Elmer SPCM-ARQH-15) for imaging and to a spectrograph with a nitrogen cooled CCD camera (Princeton Instruments SPEC-10) to record spectra.

\clearpage

\textbf{Supplementary Information} is linked to the online version of the paper at \\ www.nature.com/nature.\\

\textbf{Acknowledgements} We acknowledge P. Ruijgrok and H. van der Meer for help with the experimental setup. PZ and MO acknowledge financial support by the European Research Council (Advanced Grant SiMoSoMa). PZ acknowledges financial support by the Netherlands Organisation for Scientific Research (Veni Fellowship). PMRP acknowledges financial support by Program Ci\^{e}ncia 2008 from Funda\c{c}\~{a}o para a Ci\^{e}ncia e a Tecnologia.\\

\textbf{Author Contributions} P.Z. and M.O. designed the optical experiments, P.Z. and P.M.R.P developed the procedure for tip-functionalization, P.Z. performed the optical experiments, P.M.R.P performed the calculations, P.Z, P.M.R.P. and M.O. analyzed the data and wrote the manuscript.\\

\textbf{Author Information} Reprints and permissions information is available at \\ www.nature.com/reprints. The authors declare no competing financial interests. Correspondence and requests for materials should be addressed to M.O. (orrit@physics.leidenuniv.nl).

\clearpage

\textbf{Figure 1 Principle of the method} (a) A single gold nanorod functionalized with biotin is introduced into an environment with the protein of interest. Binding of the analyte molecules to the receptors induces a red-shift of the longitudinal SPR (exaggerated in the illustration). This shift is monitored at a single frequency using photothermal microscopy. (b) Calculation in the discrete dipole approximation of the electric field intensity around a gold nanorod, evaluated on resonance with its longitudinal SPR (supplementary information section 5). The size of the nanorod was set to 31 nm $\times$ 9 nm to match the SPR employed in the experiments. (c) Relative change in the photothermal signal (i.e. absorption cross section) as a function of the heating-laser wavelength for a red-shift of 1 nm. Plotted for SPR linewidths, $\Gamma$, of 85 meV (blue dotted line), 110 meV (red solid line), and 150 meV (green dashed line). The red square indicates the working point in our experiments, in which we use a heating laser with a wavelength of 785 nm. The calculation of the cross section was done in the electrostatic approximation. \\

\textbf{Figure 2 Photothermal time-trace showing single-molecule binding events} The normalized photothermal signal as a function of time for biotin functionalized gold nanorods in the presence of a streptavidin-R-phycoerythrin conjugate. The photothermal signal was recorded on three different nanorods in the presence of different concentrations of the protein. The red lines are fits to the time traces using a step-finding algorithm \cite{kerssemakers06}. The right-hand axis corresponds to the estimated SPR red-shift, based on a sensitivity of the photothermal signal of 5\% per nanometer red-shift, deduced from the linewidth of this particular nanorod, see Fig. \ref{principle}c. Around $t$=850s for the 0 nM case the data is masked because we checked for mechanical drift of the focus.\\

\textbf{Figure 3 Statistics of the SP shifts induced by a single molecule} (a) Histograms of the SPR red-shifts for molecules of different molecular weights. The gray bars indicate the noise level of the measurements. The vertical dashed line indicates zero SP shift. (b) Magnitude of the biggest two steps in each time trace, as a function of molecular weight of the protein. The error bars indicate the standard deviation of the population. Lines: DDA calculations of the SPR red-shift for purely refractive proteins, evaluated for two different protein conformations on the surface of the nanorod.

\clearpage

\clearpage

\clearpage

\end{document}